\documentclass[superscriptaddress,aps,prb,footnoteinbib,reprint]{revtex4-2}
\usepackage{amsmath,amssymb,amsfonts,amsthm,hyperref,longtable,graphicx}
\hypersetup{hidelinks}
\usepackage[capitalise]{cleveref}

\begin{document}
	\title{Dynamically stable negative-energy states induced by spin-transfer torques}
	\author{J. S. Harms}
	\email{j.s.harms@uu.nl}
	\affiliation{Institute for Theoretical Physics, Utrecht University, 3584CC Utrecht, The Netherlands}
	\author{A. R\"uckriegel}
	\affiliation{Institut f\"ur Theoretische Physik, Universit\"at  Frankfurt,Max-von-Laue Strasse 1, 60438 Frankfurt, Germany}
	\author{R. A. Duine}
	\affiliation{Institute for Theoretical Physics, Utrecht University, 3584CC Utrecht, The Netherlands}
	\affiliation{Department of Applied Physics, Eindhoven University of Technology, P.O. Box 513, 5600 MB Eindhoven, The Netherlands}
	\begin{abstract}
		We investigate instabilities of the magnetic ground state in ferromagnetic metals that are induced by uniform electrical currents, and, in particular, go beyond previous analyses by including dipolar interactions.
		These instabilities arise from spin-transfer torques that lead to Doppler shifted spin waves.
		For sufficiently large electrical currents, spin-wave excitations have negative energy with respect to the uniform magnetic ground state,
		while remaining dynamically stable due to dissipative spin-transfer torques.
		Hence, the uniform magnetic ground state is energetically unstable, but is not able to dynamically reach the new ground state.
		We estimate this to happen for current densities $ j\gtrsim (1-D/D_c)10^{13} \mathrm{A/m^2} $ in typical thin film experiments,
		with $ D $ the Dzyaloshinskii-Moriya interaction constant, and $ D_c $ the Dzyaloshinskii-Moriya interaction that is required for spontaneous formation of spirals or skyrmions.
		These current densities can be made arbitrarily small for ultrathin film thicknesses at the order of nanometers, due to surface- and interlayer effects. From an analogue gravity perspective, the stable negative energy states are an essential ingredient to implement event horizons for magnons -- the quanta of spin waves -- giving rise to e.g. Hawking radiation and can be used to significantly amplify spin waves in a so-called black-hole laser.
	\end{abstract}
	\maketitle

	\section{Introduction}
	Unruh's 1981 paper ''Experimental black hole evaporation''~\cite{unruh_experimental_1981} proposed that following the argument for thermal black-hole radiation~\cite{hawking_black_1974} a sonic analogue event horizon can be created by considering sound waves in a flowing medium.
	This sonic event horizon emits a thermal spectrum of sound waves and opens up possibilities for the experimental observation of Hawking radiation.
	The event horizon for sound waves is created by a transition from subsonic to supersonic background flow, such that sound waves incoming from the subsonic region cannot escape the supersonic region once they have passed the event horizon.
	Motivated by Unruh's work, theoretical proposals of analogue event horizons based on different systems were put forward~\cite{faccio_analogue_2013,novello_artificial_2002,barcelo_analogue_2005}. These include phase oscillations in a Bose-Einstein condensate~\cite{garay_sonic_2000}, slow light in dielectric media~\cite{leonhardt_space-time_2000,leonhardt_relativistic_2000}, trapped ion rings~\cite{horstmann_hawking_2010,schutzhold_analogue_2007}, Weyl semi-metals~\cite{volovik_black_2016} and, as discussed in this article, metallic magnets~\cite{roldan-molina_magnonic_2017}.
	Although Unruh's original proposal considers waves in water which can not be pushed into the quantum regime, the existence of classically stimulated Hawking emission has been observed in Ref.~\cite{weinfurtner_measurement_2011}. Furthermore, thermal Hawking radiation in a Bose-Einstein condensate, a system which might be driven to the quantum regime, has been observed in Ref.~\cite{munoz_de_nova_observation_2019}.
	
	Moreover, the combination of a black-hole and white-hole horizon -- the time-reversed partner of a black-hole horizon -- is proposed to lead to huge amplitude enhancements at specific resonant frequencies~\cite{corley_black_1999}, thereby acting as a black-hole laser. The resonance frequencies occur due to constructive interference of particle-hole coupling at each horizon, which gives rise to Hawking radiation in the quantum regime.
	An implementation of the latter is the spin-wave laser proposed in Ref.~\cite{doornenbal_spin-wave_2019},
	which provides a way of injecting spin angular momentum into a magnetic sample through amplification of spin waves,
	driven by current induced spin-transfer torques~\cite{brataas_current-induced_2012}.
	
	Spin waves are collective excitations that occur in magnetically ordered systems and correspond, at the semi-classical level, to the precession of spatially separated spins where the phase difference between them is determined by the wavelength. 
	Using spin waves for information transport and processing is the goal of magnonics~\cite{chumak_magnon_2015} -- the spin-wave analogue of electronics.
	A difficulty towards realizing spin wave based technology is the finite lifetime of spin waves resulting from processes that lead to decay of spin angular momentum. The spin-wave laser gives a potential way to compensate relaxation of spin waves by injection of spin angular momentum.
	
	In this article, we investigate energetic and dynamic instabilities of spin waves in metallic ferromagnetic thin films, induced by spin-transfer torques, i.e., torques arising from the interaction of the spin-polarized current and the magnetization dynamics~\cite{ralph_spin_2008,brataas_current-induced_2012,slonczewski_current-driven_1996,fernandez-rossier_influence_2004,duine_functional_2007,tatara_theory_2004}.
	More specifically, spin waves are Doppler shifted in the presence of an electrical current~\cite{fernandez-rossier_influence_2004,bazaliy_modification_1998}, with an effective spin-drift velocity proportional to the electrical current.
	This spin wave Doppler shift was experimentally observed by~\citet{vlaminck_current-induced_2008}.
	The spin-drift velocity, if large enough, can lead to instabilities in the ferromagnetic ground state~\cite{fernandez-rossier_influence_2004,tatara_theory_2004}.
	For the existence of analogue horizons it is important to distinguish energetic and dynamic instabilities. Energetic instabilities are characterized by the existence of negative energy excitations, while dynamical instabilities lead to exponential growth of small amplitude excitations. Contrary to most physical systems, these instabilities do not necessarily coincide for spin waves in a ferromagnetic metal, due to dissipative spin-transfer torques~\cite{ralph_spin_2008}.
	We find that magnons -- the quanta of spin waves -- can be dynamically stable for a wide range of currents that make the ferromagnetic ground state energetically unstable.
	
	In the context of analogue gravity, the magnonic event horizon is defined by the transition from a region of positive energy states to a region with dynamically stable negative energy states.
	For linearly dispersing sound waves, such as waves in water, the negative energy region corresponds to unidirectional movement of sound waves.
	In general, an event horizon is a region which couples positive energy states to negative energy states.
	For non-linearly dispersing sound waves one can still define the event horizon as the region that couples positive energy states and dynamically stable negative energy states.
	These generalized event horizons are referred to as dispersive horizons~\cite{chaline_aspects_2013}.

	The ferromagnetic thin film set-up we consider in this article is similar to~Ref.~\cite{doornenbal_spin-wave_2019}, but treated more generally, including effects of surface- and volume anisotropies, Dzyaloshinskii-Moriya interaction, dipole-dipole interactions and finite thickness of the thin film. 
	We find that the current density needed to create energetically unstable, but dynamically stable, states is of the order $ j\gtrsim (1-D/D_c)10^{13} \mathrm{A/m^2} $ for typical thin film experiments, with $ D $ the Dzyaloshinskii-Moriya constant, and $ D_c $ the Dzyaloshinskii-Moriya interaction that is required for spontaneous formation of spirals or skyrmions. 
	The critical current density can be made arbitrarily small for thin film thicknesses at the order of a nanometer.
	This decrease is primarily due to the effects of surface anisotropy and interfacial Dzyaloshinskii-Moriya interaction.
	
	The remainder of this article is organized as follows. 
	We put foreward our model and discuss spin wave solutions in~\cref{sec:magnetism_in_thin_films}.
	Furthermore, the critical current needed for energetic instabilities to exist and the region of dynamical stability are derived in~\cref{sec:energetic_and_dynamical_instabilities}.
	Additionally, we derive the critical thickness at which the ferromagnetic ground state becomes unstable due to surface and interfacial effects in~\cref{app:critical_thickness}.
	A derivation of the lowest energy dipole-exchange spin wave mode is presented in~\cref{app:dipole_exchange_mode}. We conclude with a discussion and outlook. 
	
	\section{Metallic thin film ferromagnet}\label{sec:magnetism_in_thin_films}
	\subsection{Model and set-up}
\begin{figure}
		\includegraphics[width=\columnwidth]{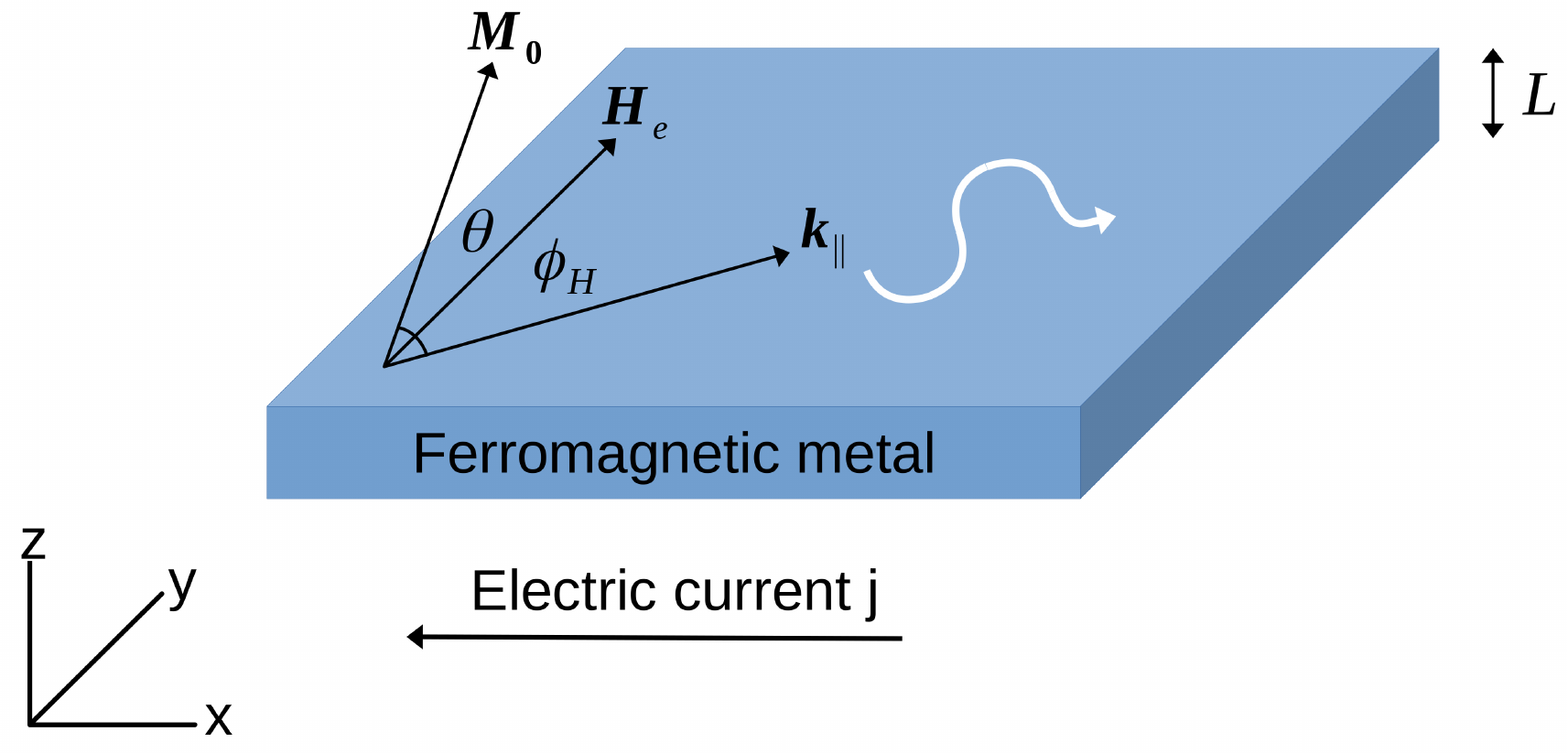}
		\caption{Sketch of the set-up. We consider a metallic ferromagnetic thin film of thickness $ L $ which is subjected to an external magnetic field $ \mathbf{H}_e $ pointing in the $ y $ direction and an electric current driven along the $ x $ direction. Furthermore, $ \theta $ is the angle of the steady state magnetisation $ \mathbf{M}_{0} $ with the plane and $ \phi_{H} $ is the angle between the spin wave propagation direction and the $ y $ axis.}
		\label{fig:set-up}
	\end{figure}

	We consider a ferromagnetic metallic thin film of thickness $ L $ in the $ z $ direction with the surfaces corresponding to $ z=\pm L/2 $. 
	We consider the set-up in~\cref{fig:set-up} that involves a thin film subject to a static external field $ \mathbf{H}_{e} $ applied in the $ y $ direction and a uniform charge current $ \mathbf{j} $ pointing in the $ -x $ direction.
	For temperatures far below the Curie temperature, amplitude fluctuations in the magnetization are negligible. In this case the dynamics of the magnetization direction $ \mathbf{n}=\mathbf{M}/M_s $ is described by the Landau-Lifschitz-Gilbert (LLG) equation, with spin-transfer torques (STTs), and Maxwell's equations in the magnetostatic limit. The LLG equation with STTs is given by~\cite{ralph_spin_2008}
	\begin{align}\label{eq:landau_lifschitz_gilbert_equation}
	\left(\partial_t+\mathbf{v}_s\cdot\nabla\right)\mathbf{n}
	&=
	-\gamma\mathbf{n}\times\mathbf{H}_{\mathrm{eff}}
	+\alpha\mathbf{n}\times\left(\partial_t+\frac{\beta}{\alpha}\mathbf{v}_s\cdot\nabla\right)\mathbf{n},
	\end{align}
	provided that spin-orbit coupling is not very strong so that spin-orbit torques are negligible. Inclusion of spin-orbit torques in our discussion is straightforward but omitted here to reduce the number of parameters. In the above equation, the adiabatic spin-transfer torque is parametrized by the velocity $ \mathbf{v}_s=-g P \mu_B\mathbf{j}/2eM_s $ that is referred to as spin-drift velocity, which is proportional to the current density $ \mathbf{j} $. Here $ g $ is the Landé factor, $ \mu_B $ the Bohr magneton, $ e $ the elementary charge, $ P $  the spin polarization of the current and $ M_s $ the saturation magnetization.
	The LLG equation describes damped precession of the magnetization around the effective field 
	$
	\mathbf{H}_{\mathrm{eff}}=-\delta E/(M_s\delta \mathbf{n}).
	$ 
	Here, $ E[\mathbf{n}] $ is the magnetic energy functional, which we consider to be of the general form
	\begin{equation}\label{eq:effective_hamiltoniaan}
	\begin{aligned}
	E=M_s\int dV &\bigg\{	
	-\frac{1}{2}J \mathbf{n}\cdot\nabla^2\mathbf{n}
	-\mu_0\mathbf{H}\cdot\mathbf{n}
	-\frac{1}{2}K_v n_z^2
	\\&
	-\frac{1}{2}D\big[\hat{y}\cdot(\mathbf{n}\times\partial_x\mathbf{n})-\hat{x}\cdot(\mathbf{n}\times\partial_y\mathbf{n})\big]
	\bigg\}.
	\end{aligned}
	\end{equation}
	In the above $ J $ is the spin stiffness, $ D $ the Dzyaloshinskii-Moriya interaction (DMI) constant that in this particular set-up may result from interfacing the magnet with a heavy metal, and $ K_v $ is the volume anisotropy constant -- this type of anisotropy is e.g. typical in the Co layer spin wave spectroscopy experiments in~Ref.~\cite{lucassen_extraction_2020}.
	The dimensionless parameters $ \alpha $ and $ \beta $ characterise the strength of the Gilbert damping parameter and the non-adiabatic spin-transfer torques, respectively.
	Usually these dissipative constants are comparable, $ \alpha \sim \beta $, and of the order $ 10^{-2} $~\cite{tserkovnyak_current_induced_dynamics}. 
	For now, we neglect surface anisotropy in the energy functional, which we discuss in~\cref{app:dipole_exchange_mode}.
	Additionally, dipole-dipole interactions are taken into account by considering the magnetostatic Maxwell's equations~\cite{damon_magnetostatic_1961}
	\begin{equation}\label{eq:magnetostatic_Maxwell_equations}
	\begin{aligned}
	\nabla\times\mathbf{H}&=\mathbf{j},\quad
	\nabla\cdot\mathbf{B}=0.
	\end{aligned}
	\end{equation}
	Here $ \mathbf{H} $ is the magnetic field strength $ \mathbf{B}=\mu_0(\mathbf{H+M}) $ the total magnetic field.
	In the steady state, the internal magnetic field $ \mathbf{H}_{0} $ and the magnetization $ \mathbf{M}_{0} $ are parallel.
	For an external magnetic field pointing in the $ y $ direction with, $ j L\ll2H_e $, the internal magnetic field and magnetization are related to the external magnetic field by	
	\begin{equation}\label{eq:magnetostatic_equilibrium}
	\begin{aligned}
		\mu_0j z\sin(\theta)
	\simeq&(\mu_0 M_s-K_v)\sin(2\theta)/2+\mu_0 H_e\sin(\theta),
	\end{aligned}
	\end{equation}
	with $ \theta $ the angle between the magnetization direction and the $ x-y $ plane.
	We find that the steady state magnetization points along the $ y $ axis if $ K_v<\mu_0(H_e+M_s-j L/2) $.
	While the steady state magnetization deviates from the $ y $ axis  if $ K_v>\mu_0(H_e+M_s-j L/2)$, where it acquires a component in the $ z $ direction.
	From this point onward we assume $ K_v<\mu_0(H_e+M_s-j L/2) $ such that the steady state magnetization is pointing in the $ y $ direction.
	Experimentally, this may be achieved by applying a sufficiently large external magnetic field.

	\subsection{Dipole-exchange spin wave modes}
%
	The dipole-exchange spin wave modes~\cite{kreisel_microscopic_2009,rezende_theory_2009,kostylev_non-reciprocity_2013,gladii_frequency_2016,lucassen_extraction_2020,de_wames_dipoleexchange_1970,wolfram_surface_1972,kalinikos_spectrum_1981,kalinikos_theory_1986} are generated by dynamical fluctuations of both the magnetization direction and the demagnetizing field, which are small compared to $ \mathbf{M}_{0}$ and $\mathbf{H}_{0} $,
	\begin{equation}\label{eq:fluctuations_magnetization_magneticfield}
	\mathbf{M}=\mathbf{M}_{0}+\mathbf{m}(t),\;\mathbf{H}=\mathbf{H}_{0}+\mathbf{h}_{\mathrm{D}}(t).
	\end{equation}
	Notice that up to linear order in the dynamical fluctuations, $ \mathbf{m} $ is perpendicular to $ \mathbf{M}_{0} $, lying in the $ x-z $ plane since we consider the magnitude of the magnetization to be constant $ |\mathbf{M}|=M_s $. Both the static and dynamic part of the magnetization and magnetic field strength should satisfy the magnetostatic Maxwell equations~(\ref{eq:magnetostatic_Maxwell_equations}). 
	We accordingly require 
	$\nabla\times\mathbf{h}_{\mathrm{D}}=0,\;\nabla\cdot\mathbf{b}=0,$
	with $ \mathbf{b}=\mu_0\left(\mathbf{h}_{\mathrm{D}}+\mathbf{m}\right) $. The first Maxwell equation allows us to write the dynamic demagnetizing field in terms of a scalar potential $ \mathbf{h}_{\mathrm{D}}=\nabla\Phi_{\mathrm{D}} $. The second Maxwell equation accordingly gives
	$
	\nabla^2\Phi_{\mathrm{D}}=-\nabla\cdot\mathbf{m},
	$
	where the magnetization $ \mathbf{m} $ outside the film is zero.
	The Landau-Lifschitz-Gilbert- and magnetostatic Maxwell equations may be rewritten by means of $ \mathbf{n}\simeq\hat{z}\,\sqrt{2}\mathrm{Re}[\Psi]+\hat{x}\,\sqrt{2}\mathrm{Im}[\Psi]+ \hat{y}\,\left(1-\left|\Psi\right|^2\right)$, with the complex field $ \Psi=(1/\sqrt{2})\left(\hat{z}+i\hat{x}\right)\cdot\mathbf{n} $. 
	In these coordinates the linearised LLG and magnetostatic Maxwell equations become
	\begin{subequations}\label{eq:linearised_equations_of_motion}
	\begin{align}
	\label{eq:linearised_LLG}
		\begin{split}
			\hat{\Omega}\Psi=&-\left(\Omega_H-\Delta_v-\Lambda^2\nabla^2\right)\Psi
		\\
		 &+ \Delta_v \Psi^* + \frac{(\partial_{z}+i\partial_{x})}{\sqrt{2}M_s}\Phi_{\mathrm{D}},
	\end{split}
\\
	\label{eq:linearised_Maxwell}	
		\frac{\nabla^2\Phi_{\mathrm{D}}}{M_s^2}=&\frac{(\partial_{z}-i\partial_{x})}{\sqrt{2}M_s}\Psi+\frac{(\partial_{z}+i\partial_{x})}{\sqrt{2}M_s}\Psi^*.
	\end{align}
	\end{subequations}
	Additionally, the exchange boundary conditions for thin films~\cite{soohoo_general_1963} require
	\begin{align}\label{eq:exchange_boundary_conditions}
	\pm\partial_z\Psi -(K_s/J)\left(\Psi+\Psi^*\right)\big\rvert_{z=\pm L/2}=0,
	\end{align}
	with $ K_s $ the surface anisotropy constant.
	In the above, we defined the following dimensionless operators and variables~\footnote{We neglected the contribution of the driving current $ \mathbf{j} $ to the magnetic field $ \mathbf{H} $ in $ \Omega_H $.}:
	dimensionless magnetic field $ \Omega_H=\mu_0H_e/\mu_0M_s, $
	dimensionless volume anisotropy $ \Delta_v=K_v/2\mu_0M_s, $
	exchange length $ \Lambda=\sqrt{J/\mu_0M_S} $
	and the dimensionless frequency operator
	$
			\hat{\Omega}=i[(1-i\alpha)\partial_t+(1-i\beta)\mathbf{v}_s\cdot\nabla+\gamma D \partial_x]/(\gamma\mu_0M_s).
	$

\begin{figure}
		\includegraphics[width=\linewidth]{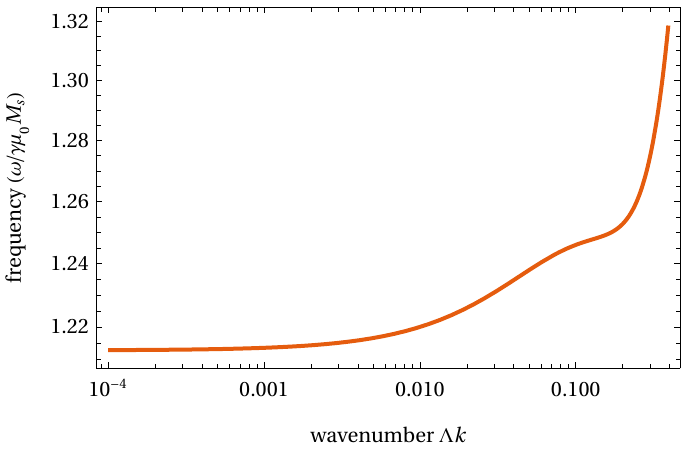}
		\caption{Dispersion relation~(\ref{eq:dispersion_real_nocurrent}) of the lowest energy spin wave mode including dipolar interactions and anisotropy. Here, we consider $ \Omega_H=1 $, $ D/\mu_0M_s=0 $, $ \Delta=-0.25 $, $ L\sim40\mathrm{nm} $ and $ \phi_H=\pi/2 $.}
		\label{fig:freqOmega}
	\end{figure}

	Using the Bogoliubov ansatz, and taking $ \mathbf{v}_s $ in the $ x $ direction, we write
	$
	\Psi(\mathbf{x},t)=u(\mathbf{x})e^{-i\omega t}+v^*(\mathbf{x})e^{i\omega^* t}
	~\text{and}~
	\Phi_\mathrm{D}(\mathbf{x},t)=w(\mathbf{x})e^{-i\omega t}+w^*(\mathbf{x})e^{i\omega^* t},
	$
	where
	$ 
	\begin{pmatrix}
		u(\mathbf{x}), &
		v(\mathbf{x}), &
		w(\mathbf{x})
	\end{pmatrix}
	\propto
	e^{i\mathbf{k}\cdot\mathbf{r}_\parallel}
	\begin{pmatrix}
		u(\mathbf{k},z), &
		v(\mathbf{k},z), &
		w(\mathbf{k},z)
	\end{pmatrix},
	$
	with
	$
	\mathbf{k}
	=
	\begin{pmatrix}
		k_x,&k_y
	\end{pmatrix}
 	$
 	and 
 	$
 	\mathbf{r}_{\parallel}
 	= 
 	\begin{pmatrix}
 		x,&y
 	\end{pmatrix}
 	$.
	The above plain wave ansatz gives rise to a spectrum of spin wave solutions.
	The lowest energy dipole-exchange spin wave dispersion relation is obtained in~\cref{app:dipole_exchange_mode} for thin films with thicknesses comparable to the exchange length $ L\sim\mathcal{O}(\Lambda) $.
	Up to linear order in $ \alpha $ and $ \beta $ the lowest energy dipole-exchange spin wave dispersion relation is given by
	 \begin{equation}
	 \begin{aligned}\label{eq:dispersion_relation_omega}
	 	(\omega_\mathbf{k}-v_sk_x)
	 	\simeq
	 	\omega^0_\mathbf{k}
	 	-i\kappa\alpha\omega^0_\mathbf{k}
	 	-i\kappa(\alpha-\beta)v_sk_x,
	 \end{aligned}
	 \end{equation}
 	 where
 	 \begin{widetext}
 	 \begin{equation}\label{eq:dispersion_real_nocurrent}
 	 	\begin{aligned}
 	 	\left(
 	 	\omega^0_{\mathbf{k}}-\gamma Dk_x
 	 	\right)/(\gamma\mu_0M_s)
 	 	=
 	 	\sqrt{
 	 	\left[\Omega_H-\Delta+\Lambda^2k^2-1/2\cos^2(\phi_{H})f(k)\right]^2
 	 	-
 	 	\left[\Delta+1/2\{1+\sin^2(\phi_{H})\}f(k)\right]^2
  		},
 	 	\end{aligned}
 	 \end{equation}
	 \end{widetext}
	 is the real part of the dispersion in the absence of an electrical current, which is plotted in~\cref{fig:freqOmega}.
	 Here, $ f(k)=1-(1-e^{-kL})/k L $ is the form factor, $ \phi_{H} $ the angle between the spin wave propagation direction and the $ y $ axis and $ \Delta\equiv\Delta_v+\Delta_s-1/2 $, with $ \Delta_s=(\Lambda/\mu_0M_sL)K_s $ the dimensionless parameter corresponding to the sum of surface anisotropies~\cite{soohoo_general_1963,gladii_frequency_2016}.
 	 In the above, $ \kappa $ is an overall factor of the imaginary part of the dispersion relation, stemming from the fact that the isotropic Gilbert damping only enters in the diagonal part of~\cref{eq:linearised_LLG}. 
 	 This term is not of importance for the stability analyses, since it remains positive in the region of interest. The precise form of $ \kappa $ can be found in~\cref{app:dipole_exchange_mode}.

	\section{Energetic and dynamical spin wave instabilities}\label{sec:energetic_and_dynamical_instabilities}
	Motivated by theoretical predictions of magnonic black/white-hole horizons~\cite{roldan-molina_magnonic_2017} and black-holes lasers~\cite{corley_black_1999,doornenbal_spin-wave_2019}, we investigate energetic and dynamic instabilities in the spin wave spectrum, due to a spin-polarized electrical current~\cite{fernandez-rossier_influence_2004}, including effects of dipole-dipole interactions, volume- and surface anisotropies, and DMI. A negative real part of the spin wave dispersion relation,~\cref{eq:dispersion_relation_omega}, indicates energetic instabilities, necessary for analogue black/white-hole setups~\cite{barcelo_analogue_2005,faccio_analogue_2013,novello_artificial_2002}.
	Dynamical instabilities on the other hand are characterized by a positive imaginary part of the spin wave dispersion relation and classically lead to an exponential growth of unstable modes.
	In contrast to most physical systems, these two types of instabilities do not necessarily coincide for the magnetization dynamics in a metallic magnetic system, due to the dissipative spin-transfer torques characterised by the parameter $ \beta $.
	\begin{figure}
		\includegraphics[width=\linewidth]{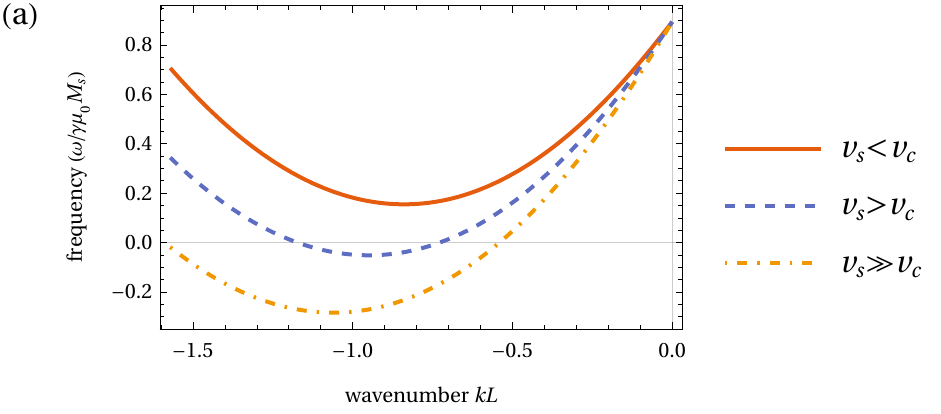}
		\includegraphics[width=\linewidth]{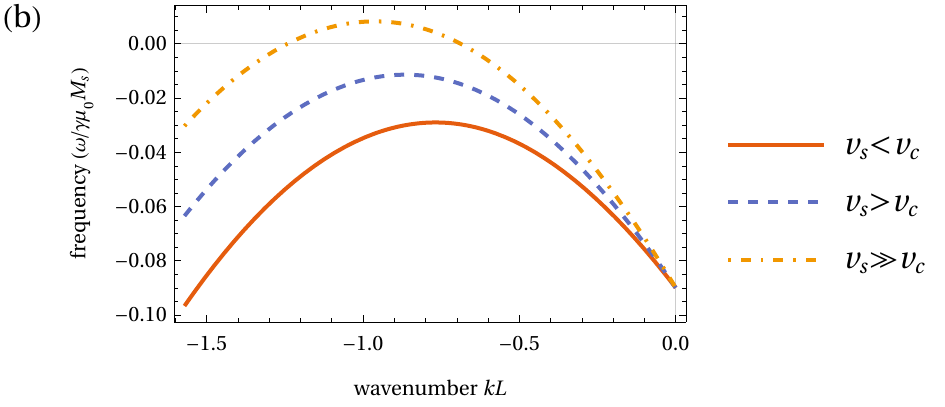}
		\caption{Dispersion relation~(\ref{eq:dispersion_relation_omega}) for different values of the spin drift-velocity $ v_s$, with $ \alpha=0.1 $ and $ \beta=0.01 $. (a) Real part of the dispersion relation. (b) Imaginary part of the dispersion relation. For small $ v_s<v_c $, spin waves are energetically and dynamically stable (red). For spin drift-velocities larger than the critical velocity, $ v_s>v_c $, we obtain energetically unstable but dynamically stable spin waves (blue). For very large $ v_s\gg v_c $ spin waves are both energetically and dynamically unstable (yellow).}
		\label{fig:instabilities}
	\end{figure}
	Accordingly, we investigate the regime in which the system is energetically unstable, but dynamically stable, see~\cref{fig:instabilities}.
	From~\cref{eq:dispersion_relation_omega} we find that the system is dynamically stable
	if
	\begin{equation}\label{eq:dynamic_instability_criterium}
		|(\alpha-\beta)v_sk_x+\alpha\gamma Dk_x|
		<
		\alpha(\gamma\mu_0 M_s)\Omega_{\mathbf{k}}
	\end{equation}
	is satisfied for all $ \mathbf{k} $, with $ \gamma\mu_0 M_s\Omega_{\mathbf{k}}=\omega^0_{\mathbf{k}}-\gamma Dk_x $ the inversion symmetric part of the dispersion relation~(\ref{eq:dispersion_real_nocurrent}).
	Energetic instabilities on the other hand are present if $ \mathrm{Re}\,\omega_{\mathbf{k}}<0 $, for some $ \mathbf{k} $. By considering minima of the dispersion relation we find the critical current above which energetic instabilities exist should satisfy $ \partial_{k_c}\mathrm{Re}\,\omega_{k_c}\rvert_{v_s=v_c}=0 $ and $ \mathrm{Re}\,\omega_{k_c}\rvert_{v_s=v_c}=0 $, for some $ k_c $. Thus, energetic instabilities are present for currents $ v_s>v_c $ and do not exist for $ v_s<v_c $, which characterises the critical velocity $ v_c $.
	The above constraints that determine the critical current are equivalent to 
	\begin{subequations}\label{eq:critical_current_constraints}
		\begin{align}\label{eq:constraint_energetic_instability}
			\partial_{k_c}\Omega_{k_c}^2&=2\Omega_{k_c}^2/k_c,
			\\ \label{eq:critical_current_constraint}
			v_c/\gamma&=[\mu_0 M_s/\sin(\phi_H)]\Omega_{k_c}/k_c-D.
	\end{align} 
	\end{subequations}
	For spin waves travelling perpendicular to the external magnetic field, $ \phi_H=\pi/2 $ in~\cref{eq:dispersion_real_nocurrent}, the constraint in~\cref{eq:constraint_energetic_instability} is explicitly written as
	\begin{equation}
		\begin{aligned}\label{eq:constraint_perpendicular_spin_wave}
			\left[\Omega_H-\Delta\right]^2
			-
			\Lambda^4k_c^{4}
			\sim
			\left[\Delta+f(k_c)\right]
			\left[\Delta+f(k_c)^2\right],
		\end{aligned}
	\end{equation}
	where we used $ f(k)-kf'(k)\sim f(k)^2 $.
	We note that $ f(k)\in[0,1] $ and typically $ \Delta=\Delta_v+\Delta_s-1/2\gtrsim-1/2 $ with $ \Omega_H\sim1 $.
	Accordingly, we assume $ [\Delta+f(k_c)][\Delta+f(k_c)^2]/[\Omega_H-\Delta]^2 $ to be small compared to unity around the critical wavelength $ k_c $.
	Next, we expand $ k_c^{2}=\kappa_{c}^{2}+\delta k_c^{2} $ around $ \Lambda^2\kappa_{c}^{2}=\Omega_H-\Delta $,
	up to linear order in $ \delta k_c^2 $ and $ [\Delta+f(k_c)][\Delta+f(k_c)^2]/[\Omega_H-\Delta]^2 $ in~\cref{eq:constraint_perpendicular_spin_wave}.
	This gives
	\begin{equation}
	\begin{aligned}
		\Lambda^2\delta k_c^{2}
		\sim&
		-\frac{1}{2}
		\frac
		{\left[\Delta+f(\kappa_{c})\right]
		\left[\Delta+f(\kappa_{c})^2\right]}
		{\Omega_H-\Delta}.
	\end{aligned}
	\end{equation}
	Similarly, we find that $ \Omega_{k_c} $, up to first order in $ \delta k_c^2 $ and $ [\Delta+f(k_c)]^2/[\Omega_H-\Delta]^2 $, is given by
	\begin{equation}
	\begin{aligned}\label{eq:perturbation_undampened_frequency}
		\Omega_{k_c}
		\sim\;&
		2[\Omega_H-\Delta]+\Lambda^2\delta k_c^{2}
		-\frac{1}{4}
		\frac
		{\left[\Delta+f(\kappa_{c})\right]^2}
		{\Omega_H-\Delta}.
	\end{aligned}
	\end{equation}
	Finally, using~\cref{eq:critical_current_constraint} we find that the critical current that generates energetic instabilities is up to linear order in $ \delta k_c^2 $ and $ [\Delta+f(k_c)]^2/[\Omega_H-\Delta]^2 $ given by
	\begin{equation}
	(v_c/\gamma\mu_0M_s)\simeq
	\left(\frac{\Omega_{k_c}}{\kappa_{c}}\right)\left(1-\frac{1}{2}\frac{\delta k_c^{2}}{\kappa_{c}^{2}}\right)-(D/\mu_0M_s).
	\end{equation}
	This can be rewritten as
	\begin{equation}\label{eq:critical_current_final}
	\begin{aligned}
	v_c=\gamma(D_c-D),
	\end{aligned}
	\end{equation}
	where
	\begin{equation}
		(D_c/2\mu_0M_s\Lambda)
		\simeq
		\sqrt[4]{
			\left[\Omega_H-\Delta\right]^2
			-(1/2)\left[\Delta+f(\kappa_c)\right]^2
		}
	\end{equation}
	is the critical DMI constant above which the ground state becomes both energetically and dynamically unstable.
	Once DMI reaches this value, the homogeneous ground state becomes unstable towards the formation of textures, typically spirals and skyrmions.
	Additionally, we note that the contribution of $ \delta k_c^2 $ drops out of the critical DMI, up to first order.

\begin{figure}
		\includegraphics[width=\linewidth]{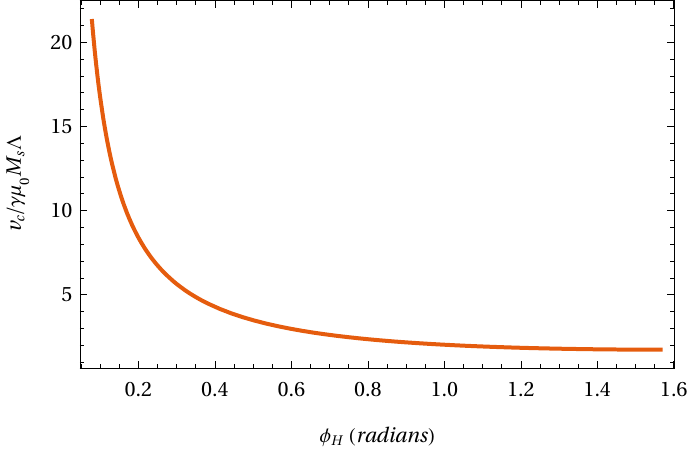}
		\caption{Numeric solution of the dimensionless critical velocity ($ v_c/\gamma\mu_0M_s\Lambda $) in~\cref{eq:critical_current_constraints} -- for dispersion relation~(\ref{eq:dispersion_real_nocurrent}) -- against the angle $ \phi_H $ in radians.
		We took the typical values $ \Omega_H\sim1 $, $ DL/\mu_0M_s\Lambda\sim 0.1 $, $ \Delta_v=K_v/2\mu_0M_s\sim 0.2  $,  $ \Delta_sL=K_s/\mu_0M_s\Lambda\sim 0.4  $ and $ L/\Lambda=3 $. }
		\label{fig:critical_current_v_angle}
	\end{figure}

	Finally, we find from~\cref{eq:dynamic_instability_criterium} that the region in which electrical currents generate energetically unstable but dynamically stable spin waves is given by
	\begin{equation}\label{eq:region_energetic_instabilities}
		\left\{
		\begin{array}{l r}
			\gamma\left(D_c-D\right)<v_s<\gamma\left(D_c-D\right)|1-\beta/\alpha|^{-1} & \beta<\alpha,\\
			\gamma\left(D_c-D\right)<v_s<\gamma\left(D_c+D\right)|1-\beta/\alpha|^{-1} & \beta>\alpha.
		\end{array}
		\right.
	\end{equation}
	This provides a large window of stability, given that usually $ \alpha\sim\beta $.
	We note that this region is determined by solely considering spin waves travelling along the $ x $ axis -- perpendicular to the external magnetic field.
	This is a consequence of the fact that the critical current for energetic and dynamic instabilities increases as spin waves travel at increasing angles $ |\phi_H-\pi/2| $ with respect to the $ x $ axis.
	In~\cref{fig:critical_current_v_angle}, we plotted the angular dependence of the critical current.
	Additionally, in the case where $ \beta/\alpha>2 $ it is possible to have dynamically unstable but energetically stable states. This occurs when the right hand side of~\cref{eq:region_energetic_instabilities} becomes smaller than the left hand side.
	As a consequence, the DMI should be at least $ D>D_c(1-2\alpha/\beta) $ for energetically unstable, but dynamically stable states to exist in the region where $ \beta/\alpha>2 $.
	Therefore, dynamically stable negative energy states are difficult to create in materials when $ \beta\gg\alpha $.
	For instance in~Ref.~\cite{chaleau_transfer_torque_2014} a value of $ \beta\sim 5\alpha $ was found.
	\begin{figure}
		\includegraphics[width=\linewidth]{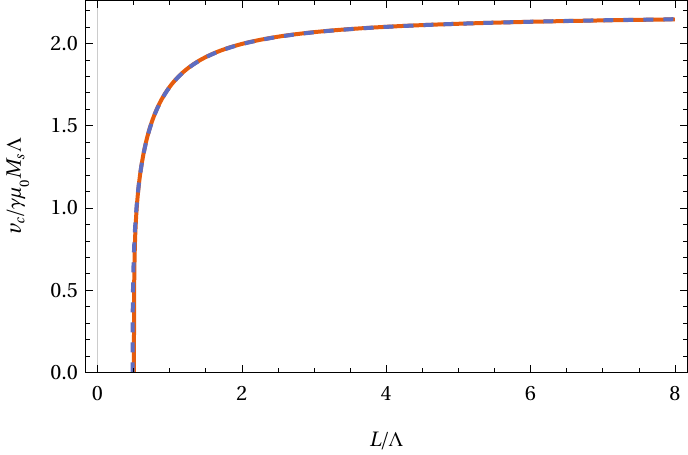}
		\caption{Dimensionless critical velocity $ v_c/\gamma\mu_0M_s\Lambda $ against dimensionless thickness $ L/\Lambda $, taking typical values $ \Omega_H\sim1 $, $ DL/\mu_0M_s\Lambda^2\sim 0.1 $, $ \Delta_v=K_v/2\mu_0M_s\sim 0.2  $,  $ \Delta_sL/\Lambda=K_s/\mu_0M_s\Lambda\sim 0.4  $. The dashed line corresponds to the linear approximation in~\cref{eq:critical_current_final} and the solid line corresponds to the numerically obtained solution of~\cref{eq:critical_current_constraints} with dispersion relation~(\ref{eq:dispersion_real_nocurrent}).}
		\label{fig:critical_current_v_thickness}
	\end{figure}

	Taking typical values for the saturation magnetization $ \mu_0M_s\sim\mu_0H_e\sim1\,\mathrm{T} $, gyromagnetic ratio $ \gamma/2\pi\sim30\mathrm{GHz\,T^{-1}} $ and exchange length $ \Lambda=\sqrt{J/\mu_0M_s}\sim 5\, \mathrm{nm} $,
	we find the typical order of magnitude of the critical current $ j_c\sim M_s|e| v_c/\mu_b \sim 10^{13}\,\mathrm{A/m^2} $, where we took $ g\sim P\sim1 $, $ M_s/\mu_B\sim 10^2\,\mathrm{nm}^{-3} $
	and $ v_c\sim\gamma\mu_0M_s\Lambda\sim10^3\,\mathrm{m/s} $.
	Furthermore, for typical values of DMI 
	$ DL/\mu_0M_s\Lambda^2\sim 0.1 $~\cite{lucassen_extraction_2020,belmeguenai_interfacial_2015,stashkevich_experimental_2015}, anisotropies $ \Delta_v=K_v/2\mu_0M_s\sim 0.2  $ and $ \Delta_sL/\Lambda=K_s/\mu_0M_s\Lambda\sim 0.4  $~\cite{lucassen_extraction_2020,gladii_frequency_2016}, we find the critical current that is needed to create energetic instabilities, given in~\cref{eq:critical_current_final}, is significantly reduced in thin films at the order of nanometers, as is shown in~\cref{fig:critical_current_v_thickness}. This reduction of the critical current is primarily due to the cumulative effect of DMI and surface anisotropies, which become prominent in ultrathin films as a consequence of their inverse scaling with respect to the thickness of the thin film.
	
	\section{Discussion and Outlook}
	We have investigated the occurrence of energetically unstable but dynamically stable spin wave excitations, due to spin-transfer torques, including effects of dipole-dipole interactions, anisotropies and DMI.
	We have shown that in typical thin film experiments~\cite{lucassen_extraction_2020,vlaminck_current-induced_2008,gladii_frequency_2016}, the critical current needed to create energetically unstable, but dynamically stable states is of the order $ j\gtrsim (1-D/D_c)10^{13} \mathrm{A/m^2} $.
	If one could experimentally enhance the DMI to be near the critical DMI, above which the homogeneous ground state becomes unstable towards the formation of textures, such as spirals and skyrmions, then a relatively small current should be sufficient to create the dynamically stable negative energy states. 
	Additionally, we found that the critical current density becomes arbitrarily small for thin film thicknesses of the order of nanometers.
	This decrease is primarily due to the cumulative effect of DMI and surface anisotropies, which become dominant in ultrathin films.
	
	Furthermore, the region in which dynamically stable negative energy spin wave excitations exist is found to be
	large, given that typically $ \alpha\sim\beta $.
	In the case where $ \beta\gg\alpha $ we note that energetically stable, dynamically unstable states could occur.
	Hence, dynamically stable negative energy states are difficult to create in materials when $ \beta\gg\alpha $.
	
	For the typical values considered in~\cref{sec:energetic_and_dynamical_instabilities}, we see a slight deviation of the first order critical velocity with respect to the numerical critical velocity at ultrathin film thicknesses,
	see~\cref{fig:critical_current_v_thickness}.
	This is due to the surface anisotropy contribution becoming larger in the ultrathin film limit,
	where the increased inaccuracy stems from the fact that we determine the critical velocity in~\cref{eq:critical_current_final} up to first order assuming $ [\Delta+f(k_c)][\Delta+f(k_c)^2] $ and $ [\Delta+f(k_c)]^2/2 $ to be small compared to $ [\Omega_H-\Delta]^2 $.
	This approximation is accurate when anisotropies are small compared to the external magnetic field, but describes the critical velocity less accurately when anisotropies become relatively large -- especially volume anisotropy -- approaching $ 2\Delta_v\lesssim\Omega_H+1 $.
	Additionally, the critical momentum $ k_c $ becomes small for ultrathin film thicknesses -- if surface anisotropies are dominating --, which makes the expansion of $ 1/k_c $ less accurate in this range.	
	When dealing with relatively large anisotropies, it is more appropriate to expand the $ k_c^2 $ around $ \kappa_c^2=\sqrt{[\Omega_H-\Delta]^2-[\Delta_v+\Delta_s]^2} $ in~\cref{eq:constraint_perpendicular_spin_wave}.
	In this case the critical DMI constant is given by
	$ (D_c/\sqrt{2}\mu_0M_s)\simeq\sqrt{(\Omega_H-\Delta+\kappa_c^2)^2-\tilde{f}(\Delta_v+\Delta_s+\tilde{f})/\kappa_c^2} $, with $ \tilde{f}=f(\kappa_c)-1/2 $.
	
	Finally, 
	energetically unstable, dynamically stable excitations are necessary to create analogue black/white-holes with spin waves~\cite{roldan-molina_magnonic_2017,faccio_analogue_2013,novello_artificial_2002,barcelo_analogue_2005}.
	Furthermore, the combination of a black- and white-hole horizon is predicted to amplify spin waves of specific frequencies, giving rise to a spin-wave laser/amplifier~\cite{corley_black_1999,doornenbal_spin-wave_2019}. 
	Future research could investigate energetic and dynamic instabilities in antiferromagnetic metals. Additionally, this model could be used to compute the resonance frequencies of the spin-wave laser in~Ref.~\cite{doornenbal_spin-wave_2019} more realistically. Moreover, one could investigate non-linear effects in such a setup, since non-linear effects quickly become important around the resonance frequencies.
	The non-linear regime could be investigated by means of the stochastic Landau-Lifschitz-Gilbert equation.
	\begin{acknowledgements}
		This work is part of the  research  programme  Fluid  Spintronics  with  projectnumber 182.069, which is (partly) financed by the Dutch Research Council (NWO).
		R.D. is member of the D-ITP consortium, a program
		of the Dutch Organization for Scientific Research (NWO)
		that is funded by the Dutch Ministry of Education, Cul-
		ture and Science (OCW).
	\end{acknowledgements}

	\appendix
	\section{Critical thickness of energetic instabilities at zero current}\label{app:critical_thickness}
	In this Appendix, we determine the critical thickness at which spin wave excitations become energetically unstable at zero electrical current.
	These energetic instabilities are due the increase in magnitude of surface anisotropies and DMI in the ultra thin film limit and are dynamically unstable by~\cref{eq:dynamic_instability_criterium}. Additionally, the range of electrical currents that generate energetically unstable but dynamically stable spin wave excitations decreases when approaching the critical thickness. This is a direct consequence of decreasing the critical current, see~\cref{sec:energetic_and_dynamical_instabilities}. 
	If surface anisotropies are dominant at small thicknesses, the critical thickness at which instabilities appear, at zero current and vanishing DMI, may be approximated at zero'th order by closing the spin-wave gap in~\cref{eq:dispersion_real_nocurrent}, giving 
	\begin{equation}\label{eq:critical_thickness_spin_wave_gap}
	\Omega_H-2\Delta_0\equiv
	\Omega_H+1-2\Delta_v-2\Delta_s^*/L_0\sim0,
	\end{equation}
	with $ k_cL\ll1 $ and $ \Delta_s^*=\Delta_s L $ constant.
	For non-vanishing DMI, up to first order in $ f(\kappa_c)\sim\kappa_cL/2 \ll1$, we find that~\cref{eq:critical_current_constraints} is equivalent to
	\begin{align}
		\label{eq:constraint_thickness_1}
		\left[\Omega_H-\Delta\right]^2
		&\sim
		\left[\Delta+f(k_c)\right]\Delta,
		\\\label{eq:constraint_thickness_2}
		{k_c}\big[
		2\left(\Omega_H-\Delta\right)
		-(D^*/L)^2
		\big]
		&\sim
		\left[\Delta+f(k_c)\right]f'(k_c),
	\end{align}
	where $ D^*=(D/\mu_0 M_s\Lambda)L$ is constant.
	We expand the above equations around $ \Delta=\Delta_0+\delta\Delta $, with $ \delta \Delta=-(\Delta_s^*/L_0^2)\delta L+\mathcal{O}(\delta L^2)$.
	Up to first order~\cref{eq:constraint_thickness_1} gives
	\begin{equation}
	k_cL_0/2\sim-4\delta\Delta.
	\end{equation}
	We find that~\cref{eq:constraint_thickness_2} in combination with the above equation leads to
	\begin{equation}
	\begin{aligned}
		\delta L
		\sim
		\frac{\Omega_H L_0^2}{32\Delta_s^*}
		\Bigg/
		\Bigg\{&
		\Omega_H\left[\frac{1}{L_0^2}+\frac{1}{6}-\frac{L_0}{32\Delta_s^*}\right]
		\\&
		-\frac{3}{16}
		-\left(\frac{D^*}{L_0^2}\right)^2
		\Bigg\},
	\end{aligned}
	\end{equation}
	with $ L_0=2\Delta_s^*/(2\Omega_H+1-2\Delta_v) $ given by~\cref{eq:critical_thickness_spin_wave_gap}.
	We thus find that the critical thickness for spin wave instabilities is given by $ L\sim L_0+\delta L $, in the case that surface anisotropies dominate in the ultrathin film.
	\section{Approximate dipole-exchange mode in thin films}\label{app:dipole_exchange_mode}	
	Here, we discuss an analytic approximation of the lowest energy spin wave dispersion relation for the setup discussed in~\cref{sec:magnetism_in_thin_films}, using the thin film magnetostatic Greens function~\cite{guslienko_magnetostatic_2011,kalinikos_spectrum_1981,kalinikos_theory_1986,kalinikos_spectrum_1981,kostylev_non-reciprocity_2013,gladii_frequency_2016}.
	We start by expressing the demagnetizing field $ \mathbf{h}_{\mathrm{D}} $ in~\cref{eq:fluctuations_magnetization_magneticfield,eq:linearised_LLG,eq:linearised_Maxwell} in terms of the magnetization by using the magnetostatic Greens function.
	It is explicitly given by
	\begin{equation}\label{eq:linearised_demagnetization_field}
		\mathbf{h}_{\mathrm{D}}(\mathbf{k},z)=\int_{-L/2}^{L/2}dz'\mathbf{G}(\mathbf{k};z-z')\mathbf{m}(\mathbf{k};z').
	\end{equation}
	Where the magnetostatic Greens function~\cite{kalinikos_spectrum_1981,guslienko_magnetostatic_2011} satisfies the magnetostatic Maxwell equation~\cref{eq:linearised_Maxwell} along with the appropriate boundary conditions~\cite{jackson_classical_1998}.
	It is explicitly given by,
	\begin{equation}
		G_{\hat{\zeta}\hat{\eta}\hat{z}}(k;z,z')
		=
		\begin{pmatrix}
			-G_{p}&0&iG_{q}\\
			0&0&0\\
			iG_{q}&0&G_{p}-\delta(z-z')
		\end{pmatrix},
	\end{equation}
	with $ \hat{\zeta}\propto\mathbf{k} $ the in-plane direction of  spin wave propagation, $ \hat{\eta} $ the orthogonal in-plane direction, $ \hat{z} $ the thickness direction,
	$
	G_{p}=\frac{|k|}{2}\exp(-|k||z-z'|) 
	$
	and
	$
	G_{q}=G_p\,\mathrm{sign}(z-z').
	$
	By substituting~\cref{eq:linearised_demagnetization_field} into the linearised Landau-Lifschitz-Gilbert equation~\cref{eq:linearised_LLG} we obtain the effective linearised LLG equation
	\begin{equation}\label{eq:LLG_Greens_function}
		\begin{aligned}
			\hat{\Omega}\Psi=&-\left[\Omega_H-\Delta_{v/s}-\Lambda^2\nabla^2\right]\Psi + \Delta_{v/s} \Psi^*
			\\
			&+
			\int_{-L/2}^{L/2}\frac{1}{2}
			\Big[
			A_k \Psi_k
			+
			B_k \Psi^*_{-k}
			\Big]dz',
		\end{aligned}
	\end{equation}
	where
	$
	A_k
	=\cos^2(\phi_H)G_p-\delta(z-z'),
	~
	B_k
	=\{1+\sin^2(\phi_H)\}G_p-2\sin(\phi_H)G_q-\delta(z-z')
	$
	and
	$ \phi_H $ the angle between wave vector $ \mathbf{k} $ and the $ y $ axis.
	Additionally we introduced the dimensionless anisotropy constant $ \Delta_{v/s}=K_{v/s}/2\mu_0M_s $, where we took $
	K_{v/s}=K_v+K_s^-\delta(z-L/2)+K_s^+\delta(z+L/2) $.
	Here $ K_s $ correspond to the surface anisotropies of the thin film.
	Following~\citet{gladii_frequency_2016} we add the term $ (K_s^{\pm}/2)\delta(z\pm L/2)n_z^2 $ in the energy functional~\cref{eq:effective_hamiltoniaan} to account for surface anisotropies.
	This differs from the approach used by~\citet{kalinikos_theory_1986} where surface anisotropies determine the exchange boundary conditions of the thin film~\cite{soohoo_general_1963}.

	Using the Bogoliubov ansatz 
	$
	\Psi(\mathbf{x},t)=u(\mathbf{x})e^{-i\omega t}+v^*(\mathbf{x})e^{i\omega^* t},
	$
	with
	$
	\begin{pmatrix}
		u(\mathbf{x}),&
		v(\mathbf{x})
	\end{pmatrix}
	=
	\int\frac{d^2\mathbf{k}}{2\pi}e^{i\mathbf{k}\cdot\mathbf{r}_\parallel}
	\begin{pmatrix}
		u(\mathbf{k},z),&
		v(\mathbf{k},z)
	\end{pmatrix},
	$
	the linearised LLG equation becomes
	\begin{equation}\label{eq:LLG_Greens_Function_Bogoliubov}
	\begin{aligned}
		\begin{pmatrix}
			\hat{F}+1/2 & 1/2-\Delta_{v/s}\\
			\Delta_{v/s}-1/2 & -\hat{F}^*-1/2
		\end{pmatrix}
		\begin{pmatrix}
			u(k,z)\\
			v(k,z)
		\end{pmatrix}&\\
		+\int_{-L/2}^{L/2}dz'
		\begin{pmatrix}
			- C(s) & D^+(s)\\
			-D^-(s) & C(s)
		\end{pmatrix}
		\begin{pmatrix}
			u(k,z')\\
			v(k,z')
		\end{pmatrix}
		&=0,
	\end{aligned}
	\end{equation}
	with $ s=z-z' $,
	$ \hat{F}=\Omega+\left(\Omega_H-\Delta_{v}+\Lambda^2k^2-\Lambda^2\partial_z^2\right) $, 
	$ \hat{F}^*=-\Omega^*+\left(\Omega_H-\Delta_{v}+\Lambda^2k^2-\Lambda^2\partial_z^2\right) $,
	$ C(s)=(1/2)\cos^2(\phi_\zeta)G_p $
	and
	$ D^\pm(s)=-(1/2)\{1+\sin^2(\phi_\zeta)\}G_p\pm|\sin(\phi_\zeta)|G_q. $
	Furthermore,
	$
	\gamma\mu_0M_s\Omega
	=
	(1+i\alpha)\omega+(1+i\beta)v_sk_x+\gamma D k_x
	$
	and
	$ 
	\gamma\mu_0M_s\Omega^*
	=
	(1-i\alpha)\omega+(1-i\beta)v_sk_x+\gamma D k_x.
	$
	
	The magnetization profile in the thickness direction may be expanded in eigenfunctions of the unpinned exchange boundary conditions, which form a complete basis~\cite{kalinikos_spectrum_1981}.
	We approximate the magnetization profile of the lowest mode by the lowest Fourier mode, for thicknesses of the order 
	$ L\sim\mathcal{O}(\Lambda) $,  
	\begin{equation}\label{eq:Fourier_expansion_thickness_direction}
		\begin{pmatrix}
			u(k,z)\\
			v(k,z)
		\end{pmatrix}
		\sim
		u^0(k)
		\sqrt{\frac{1}{L}}
		\begin{pmatrix}
			1\\
			0
		\end{pmatrix}
		+
		v^0(k)
		\sqrt{\frac{1}{L}}
		\begin{pmatrix}
			0\\
			1
		\end{pmatrix},
	\end{equation}
	which is the uniform mode approximation.
	Using the above ansatz the linearised LLG equation~\cref{eq:LLG_Greens_Function_Bogoliubov} becomes
	\begin{equation}
		\begin{pmatrix}
			\Omega+\Omega_{\mathrm{d}}&-\Omega_{\mathrm{i}}\\
			\Omega_{\mathrm{i}}&\Omega^*-\Omega_{\mathrm{d}}\\
		\end{pmatrix}
		=0,
	\end{equation}
	where
	\begin{subequations}
		\begin{align}
			\Omega_{\mathrm{d}}
			&=
			\Omega_H-\Delta+\Lambda^2k^2-1/2\cos^2(\phi_H)f(k),\\
			\Omega_{\mathrm{i}}
			&=
			\Delta+1/2\{1+\sin^2(\phi_H)\}f(k).
		\end{align}
	\end{subequations}
	With $ f(k)=1-(1-e^{-kL})/kL $, $ \Delta=\Delta_{v}+\Delta_s-1/2 $ and $ \Delta_s=(\Lambda/2\mu_0M_sL)\left(K_s^-+K_s^+\right) $.
	Hence, the lowest mode dispersion relation, up to first order in $ \alpha $ and $ \beta $, is given by
	\begin{equation}
		\begin{aligned}
			(\omega_\mathbf{k}-v_sk_x)
			\simeq
			\omega^0_\mathbf{k}
			-i\kappa\alpha\omega^0_\mathbf{k}
			-i\kappa(\alpha-\beta)v_sk_x,
		\end{aligned}
	\end{equation}
	with $ \left(\omega^0_{\mathbf{k}}-\gamma Dk_x\right)^2/(\gamma\mu_0M_s)^2=\Omega_\mathbf{k}^2 $, $ \kappa=(\Omega_d/\Omega_\mathbf{k}) $ and
	 \begin{align}
	 \begin{aligned}
	 	\Omega_{\mathbf{k}}^2
		=
		\left[\Omega_H-\Delta+\Lambda^2k^2-1/2\cos^2(\phi_{H})f(k)\right]&^2
		\\
		-\left[\Delta+1/2\{1+\sin^2(\phi_{H})\}f(k)\right]&^2.
	 \end{aligned}
	\end{align}
	\bibliographystyle{apsrev4-2}
	\bibliography{./bibliographyenergeticstability}
	\end{document}